# Grand challenges in altmetrics: heterogeneity, data quality and dependencies

Stefanie Haustein

**Abstract**

As uptake among researchers is constantly increasing, social media are finding their way into scholarly communication and, under the umbrella term altmetrics, were introduced to research evaluation. Fueled by technological possibilities and an increasing demand to demonstrate impact beyond the scientific community, altmetrics received great attention as potential democratizers of the scientific reward system and indicators of societal impact. This paper focuses on current challenges of altmetrics. Heterogeneity, data quality and particular dependencies are identified as the three major issues and discussed in detail with a particular emphasis on past developments in bibliometrics. The heterogeneity of altmetrics mirrors the diversity of the types of underlying acts, most of which take place on social media platforms. This heterogeneity has made it difficult to establish a common definition or conceptual framework. Data quality issues become apparent in the lack of accuracy, consistency and replicability of various altmetrics, which is largely affected by the dynamic nature of social media events. It is further highlighted that altmetrics are shaped by technical possibilities and depend particularly on the availability of APIs and DOIs, are strongly dependent on data providers and aggregators, and potentially influenced by technical affordances of underlying platforms.

## 1  Introduction

Social media have profoundly changed how people communicate. They are now finding their way into scholarly communication, as researchers increasingly use them to raise their visibility, connect with others and diffuse their work (Rowlands, Nicholas, Russell, Canty, & Watkinson, 2011; Van Noorden, 2014). Scholarly communication itself has remained relatively stable; in the course of its 350-year history the scientific journal has not altered much. Even in the digital age, which has facilitated collaboration and increased the speed of publishing, the electronic journal article remains in essence identical to its print counterpart. Today, the peer-reviewed scientific journal still is the most important channel to diffuse scientific knowledge.

In the context of the diversification of the scholarly communication process brought about by the digital era, social media is believed to increase transparency: ideas and results can be openly discussed and scrutinized in blog posts, some journals and designated platforms are making the peer-review process visible, data and software code are increasingly published online and reused, and manuscripts and presentations are being shared on social media. This diversification of the scholarly communication process presents both an opportunity and a challenge to the scholarly community. On the one hand, researchers are able to distribute various types of scholarly work and reach larger audiences; on the other hand, this leads to a further increase of information overload. At first, altmetrics were seen as an improved filter to overcome the information overload stemming from the diversification and increase in scholarly outputs (Priem, Taraborelli, Groth, & Neylon, 2010). In that sense, quite a few parallels exist between the development of bibliometrics and altmetrics:

> It is too much to expect a research worker to spend an inordinate amount of time searching for the bibliographic descendants of antecedent papers. It would not be excessive to demand that the thorough scholar check all papers that

> have cited or criticized such papers, if they could be located quickly. The citation index makes this check practicable. (Garfield, 1955, p. 108)

> No one can read everything. We rely on filters to make sense of the scholarly literature, but the narrow, traditional filters are being swamped. However, the growth of new, online scholarly tools allows us to make new filters; these altmetrics reflect the broad, rapid impact of scholarship in this burgeoning ecosystem. (Priem et al., 2010, para. 1)

While altmetrics rely on users of various social media platforms to identify the most relevant publications, datasets and findings, Garfield (1955) and before him Gross and Gross (1927) believed that citing authors would outperform professional indexers in identifying the most relevant journals, papers and ideas. Both altmetrics and citation indexing thus rely on collective intelligence, or the wisdom of the crowds, to identify the most relevant scholarly works. Citation indexing can, in fact, be described as an early, pre-social web version of *crowdsourcing*. Quite similarly to the reoccurring question about the meaning of altmetrics, early citation analysts admitted that they did "not yet have any clear idea about what exactly [they] were] measuring" (Gilbert, 1977, p. 114) and the interpretation of citations as indicators of impact remains disputed from a social constructivist perspective on the act of citing. However, the fundamental difference between indicators based on citations and social media activity is that the act of citing has been an essential part of scholarly communication in modern science, while researchers are still exploring how to use social media.

Both bibliometrics and altmetrics share the same fate of being (too) quickly identified and used as indicators of impact and scientific performance. Although early applications of bibliometric indicators in research management emphasized their complementary nature and the need for improving, verifying and triangulating available data through experts (Moed, Burger, Frankfort, & Van Raan, 1985), citations soon became a synonym for scientific impact and quality. Consequently, bibliometric indicators were misused in university and journal rankings as well as in individual hiring and promotion decisions, which has, in turn, led to adverse effects such as salami publishing and self-plagiarism, honorary authorship, authorship for sale, as well as strategic citing through self-citation or citation cartels (for a recent overview compare Haustein & Larivière, 2015). Even though altmetrics are presented as a way to counter-balance the obsession with and influence of indicators such as the impact factor or h-index, and make research evaluation fairer by considering more diverse types of scholarly works and impact (Piwowar, 2013; Priem & Hemminger, 2010), they run the risk of causing similar effects, particularly as they have been created in the midst of a technological push and a policy pull. This paper focuses on the current challenges of altmetrics; a particular emphasis is placed on how the current development of this new family of indicators compares with the past development of bibliometrics.

## 2 Challenges of altmetrics

Altmetrics face as many challenges as they offer opportunities. In the following section, three major—what the author considers the most profound—challenges are identified and discussed. These include the heterogeneity, data quality issues and specific dependencies of altmetrics.

### 2.1 Heterogeneity

The main opportunity provided by altmetrics—their variety or heterogeneity—represents also one of their major challenges. Altmetrics comprise many different types of metrics, which has made it difficult to establish a clear-cut definition of what they represent. The fact that they have been considered as a unified, monolithic alternative to citations has hindered discussions, definitions, and interpretations of what they actually measure: why would things as diverse as a mention on Twitter, an expert recommendation on F1000Prime, a reader count on Mendeley, a like on Facebook, a citation in a blog post and the reuse of a dataset share a common meaning? And why are they supposed to be inherently different from a citation, which can itself occur in various shapes: a perfunctory mention in the introduction, a direct quote to highlight a specific argument, or a reference to acknowledge the reuse of a method? In the following, the challenges associated with their heterogeneity and lack of meaning are discussed by addressing the absence of a common definition, the variety of social media acts, users and their motivation, as well as the lack of a conceptual framework or theory.

**Lack of a common definition**

Although altmetrics are commonly understood as online metrics that measure scholarly impact alternatively to traditional citations, a clear definition of altmetrics is lacking. Priem (2014, p. 266) rather broadly defined altmetrics as the "study and use of scholarly impact measures based on activity in online tools and environments", while the altmetrics manifesto refers to them as elements of online traces of scholarly impact (Priem et al., 2010), a definition that is similar to *webometrics* (Björneborn & Ingwersen, 2004) and congruent with the *polymorphous mentions* described by Cronin, Snyder, Rosenbaum, Martinson, and Callahan (1998). Moed (2016, p. 362) conceptualizes altmetrics as "traces of the computerization of the research process". A less abstract definition of what constitutes an altmetric is, however, absent and varies between authors, publishers, and altmetrics aggregators. As it was shown that these new metrics are rather complementary than an alternative to citation-based indicators, the term has been criticized and suggested to be replaced by *influmetrics* (Rousseau & Ye, 2013) or *social media metrics* (Haustein, Larivière, Thelwall, Amyot, & Peters, 2014). There is also much confusion between altmetrics and article level metrics (ALM), which as a level of aggregation can refer to any type of metric aggregated for articles. While metrics based on social media represent the core of altmetrics, some also consider news media, policy documents, library holdings and download statistics as relevant sources, although derived indicators for those have been available long before the introduction of altmetrics (Glänzel & Gorraiz, 2015). Despite often being presented as antagonistic, some of these metrics are actually similar to journal citations (e.g., mention in a blog post), while others are quite different (e.g., tweets). As a consequence, various altmetrics can be located on either side of citations on a spectrum from low to high levels of engagement with scholarly content (Haustein, Bowman, & Costas, 2016). Moreover, the landscape of altmetrics is constantly changing. The challenge of the lack of a common definition can thus be only overcome if altmetrics are integrated into one metrics toolbox:

> It may be time to stop labeling these terms as parallel and oppositional (i.e., altmetrics vs bibliometrics) and instead think of all of them as available scholarly metrics—with varying validity depending on context and function. (Haustein, Sugimoto, & Larivière, 2015, p. 3)

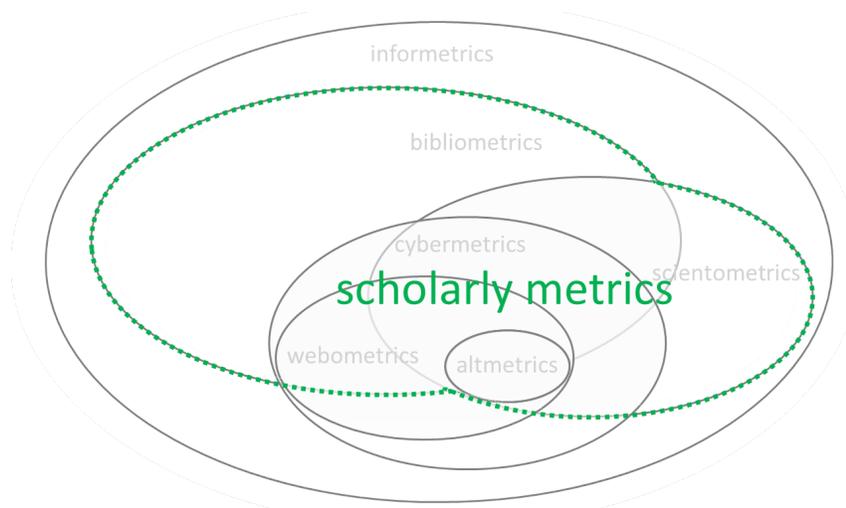

**Figure 1.** The definition of scholarly metrics and the position of altmetrics in informetrics, adapted from Björneborn and Ingwersen (2004, p. 1217). Sizes of the ellipses are not representative of the size fields but made for the sake of clarity only.

Following Haustein, Sugimoto and Larivière (2015) and employing the terminology and framework used by Haustein, Bowman and Costas (2016) scholarly metrics are thus defined as indicators based on recorded events of acts (e.g., viewing, reading, saving, diffusing, mentioning, citing, reusing, modifying) related to scholarly documents (e.g., papers, books, blog posts, datasets, code) or scholarly agents (e.g., researchers, universities, funders, journals). Hence, altmetrics refer to a heterogeneous subset of scholarly metrics and are a proper subset of informetrics, scientometrics and webometrics (Fig. 1).

**Heterogeneity of social media acts, users and motivations**

As shown above, "altmetrics are indeed representing very different things" (Lin & Fenner, 2013, p. 20). Even if one considers only those based on social media activity, altmetrics comprise anything from quick—sometimes even automated—mentions in microposts to elaborate discussions in expert recommendations. The diversity of indicators is caused by the different nature of platforms, which entail diverse user populations and motivations. This, in turn, affects the meaning of the derived indicator. It is thus futile to speak about the *one* meaning of altmetrics but rather the meaning of specific types or groups of metrics. Although many platforms now incorporate several functions, which aggravates classification, the following seven groups of social media platforms used for altmetrics are identified:

   a. social networking (e.g., Facebook, ResearchGate)
   b. social bookmarking and reference management (e.g., Mendeley, Zotero)
   c. social data sharing including sharing of datasets, software code, presentations, figures and videos, etc. (e.g., Figshare, Github)
   d. blogging (e.g., ResearchBlogging, Wordpress)
   e. microblogging (e.g., Twitter, Weibo)
   f. wikis (e.g., Wikipedia)
   g. social recommending, rating and reviewing (e.g., Reddit, F1000Prime)

The different purposes and functionalities of these platforms attract different audiences to perform various kinds of acts. For example, recommending a paper is inherently different from saving it to a reference manager and blogging about a dataset differs from tweeting it (Taylor, 2013). These differences are reflected in the metrics derived from these platforms. For example, the selectivity and high level of engagement associated with blogging and recommending is mirrored in the low percentage of papers linked to these events. Less than 2% of recent publications get mentioned in blog posts and the percentage of papers being recommended on F1000Prime is equally sparse. On the other hand, Twitter and Mendeley coverage is much higher at around 10-20% and 60-80% (Haustein, Costas, & Larivière, 2015; Priem, Piwowar, & Hemminger, 2012; Thelwall & Wilson, 2015; Waltman & Costas, 2014). This reflects both the lower level of engagement of tweeting and saving to Mendeley compared to writing a blog post or an F1000Prime review, as well as the size of the Twitter and Mendeley user populations in comparison to the number of bloggers and F1000Prime experts. Although the signal for Mendeley is quite high at the paper level, the uptake of the platform among researchers is rather low (6-8%). Most surveys report that around 10-15% of researchers use Twitter for professional purposes, while ResearchGate, LinkedIn and Facebook were more popular although passive use prevailed (Mas-Bleda, Thelwall, Kousha, & Aguillo, 2014; Rowlands et al., 2011; Van Noorden, 2014). The different types of uses and compositions of users echo in correlations with citations. Moderate to high positive correlations were found between citations and Mendeley reader counts and F1000Prime recommendations (Bornmann & Leydesdorff, 2013; Li, Thelwall, & Giustini, 2012; Thelwall & Wilson, 2015), which alludes to the academic use and users of these platforms. F1000Prime 'faculty members' are a selected group of experts in their fields, while the majority of the Mendeley users have been shown to consist of students and early-career researchers (Zahedi, Costas, & Wouters, 2014). On the other hand, correlations between citations and tweets are rather weak, which suggests non-academic use (Costas, Zahedi, & Wouters, 2015; Haustein, Costas, et al., 2015). More detailed information about user demographics and particularly their motivation to interact with scholarly contents on social media is, however, still mostly lacking.

The heterogeneity of altmetrics is not only apparent between the seven types of sources but also within. For example, Facebook differs from ResearchGate and Reddit from F1000Prime. Even the exact same event can be diverse, for example, tweets can comprise the promotion of one's own work (self-tweets), the diffusion of a relevant paper, the appraisal of a method or the criticism of results. A Mendeley readership count can imply a quick look at or intense reading of a publication. Moreover, not all recorded events—e.g., a tweet linking to a paper—are based on direct acts, but could be automatically created (Haustein, Bowman, Holmberg, et al., 2016), and not all acts result in a recorded event, if, for example, the absence of identifiers to sources prevents capturing them.

**Lack of conceptual frameworks and theories**

The lack of a theoretical foundation coupled with its pure data-drivenness is a central and reoccurring criticism of altmetrics. What constitutes an altmetric today is almost entirely determined by the availability of data and the ease with which it can be collected. To be fair, after decades of debates the field of scientometrics has not been successful in implementing an overarching theory either (Sugimoto, 2016). And with the creation of the Science Citation Index, the field of bibliometrics and the use of citation analysis in research evaluation have been driven by the availability of data to the point that it has come to shape the definition of scientific impact. Despite being "inherently subjective, motivationally messy and susceptible to abuse" (Cronin, 2016, p. 13), the act of citing has been a central part of the scholarly communication process since the early days of modern science, while social media is still searching for its place in academia. In the context of altmetrics, the lack of a theoretical scaffold is caused by the sudden ease of data collection and the high demand by funding bodies to prove societal impact of research. It is further impeded by the previously described heterogeneity of metrics. Based on a framework of various acts, which *access*, *appraise* or *apply* scholarly documents or agents, Haustein, Bowman and Costas (2016) discussed different altmetrics in the light of normative and social constructivist theories, the concept symbols citation theory as well as social capital, attention economics and impression management. These theories were selected due to the affinity of altmetrics with citations, as well as the imminent social aspect of social media. The fact that the different theories apply better to some of these acts, and are less suitable for others, further emphasizes the heterogeneity of altmetrics. Haustein, Bowman and Costas (2016) demonstrated that the discussed theories help to interpret the meaning of altmetrics but stress that they are not able to fully explain acts on social media. Other theories as well as more qualitative and quantitative research are needed to understand and interpret altmetrics.

## 2.2 Data quality

The central importance of data quality cannot be emphasized enough, particularly in the context of research evaluation. Data quality, including metadata quality and errors, as well as coverage have thus been a reoccurring topic in bibliometrics, with a focus on comparing the three major citation indexes Web of Science, Scopus and Google Scholar. In altmetrics, data quality is a major challenge and transcends the known errors and biases for citation data. In the context of citations, errors mostly represent discrepancies between the act and the recorded event. These can be discovered and measured either through triangulation of different data aggregators (i.e., citation indexes) or by referring to the original source (i.e., the reference list in a publication). While bibliometrics sources are static documents, most data sources in the context of altmetrics are dynamic, which can be altered or deleted entirely. Accuracy, consistency and replicability can be identified as the main issues of altmetrics data quality. Potential data quality issues can occur at the level of data providers (e.g., Mendeley, Twitter), data aggregators (e.g., Altmetric, ImpactStory, Plum Analytics) and users. It should be noted that many providers (e.g., Twitter, Facebook, Reddit) are not targeted at academia and altmetrics data quality is thus not their priority. In the context of the NISO Altmetrics Initiative[1], a working group on altmetrics data quality[2] is currently drafting a code of conduct to identify and improve data quality issues.

Research into altmetrics data quality is still preliminary and has so far mainly focused on the accuracy of Mendeley data (Zahedi, Bowman, & Haustein, 2014) and the consistency between altmetrics aggregators (Jobmann et al., 2014; Zahedi, Fenner, & Costas, 2015, 2014). Inconsistencies between data aggregators can be partly explained by different retrieval strategies. For example, Lagotto aggregates Facebook shares, likes and comments, while Altmetric records only public posts. Recording tweets continuously and in real time, Altmetric shows the highest coverage of papers on Twitter, while Lagotto captures only a fraction (Zahedi et al., 2015). The extent to which Altmetric's record of tweets to scientific papers is accurate or complete is unknown. Replication would only be possible through a direct verification against Twitter's data, which is precluded by the costliness of access. The replicability of altmetrics is further impeded by the dynamic nature of events. While citations cannot decrease because they cannot be deleted, Mendeley readership counts can change over time (Bar-Ilan, 2014). The provision of timestamps for events and longitudinal statistics might be able to mitigate replicability issues at least to some extent.

---

[1] http://www.niso.org/topics/tl/altmetrics_initiative/
[2] http://www.niso.org/publications/newsline/2015/working_group_connection_apr2015.html#bi0

The quality of altmetrics data is also affected by the metadata of scholarly works. One central aspect is the availability of certain identifiers (particularly the DOI), which is discussed below in the context of dependencies. Moreover, the metadata of traditional publications might not be sufficient to construct and (potentially) normalize altmetric indicators in a meaningful manner. The publication year, which is the time-stamp used for citation indicators, is not appropriate for social media events, which happen within hours of publication and exhibit half-lives that can be counted in days rather than years (Haustein, Bowman, & Costas, 2015). Whether or not to aggregate altmetrics for various versions of documents (e.g., preprint, version of record) is also a central issue of debate. While the location—that is, the journal website, repository or author's homepage—of a document makes hardly any difference for citations, it is essential for tracking related events on social media.

## 2.3 Dependencies

Altmetrics have been and continue to develop under the pressure of various stakeholders. They are explicitly driven by technology and represent a *computerization movement* (Moed, 2016). The availability of big data and the ease with which they can be assessed was met by a growing demand, particularly by research funders and managers, to make the societal impact of science measurable (Dinsmore, Allen, & Dolby, 2014; Higher Education Funding Council for England, 2011; Wilsdon et al., 2015)—despite the current lack of evidence that social media events can serve as appropriate indicators of societal impact. Along these lines, the role of publishers in the development of altmetrics needs to be emphasized. Owned by large for-profit publishers, Altmetric, Plum Analytics and Mendeley operate under a certain pressure to highlight the value of altmetrics and to make them profitable. Similarly, many journals have started to implement altmetrics, not least as a marketing instrument.

The majority of collected altmetrics are those that are comparably easily captured, often with the help of APIs and document identifiers. Activity on ResearchGate or Zotero is, for example, not used because these platforms do not (yet) offer APIs. The strong reliance on identifiers such as DOIs creates particular biases against fields and countries where they are not commonly used, such as the social sciences and humanities (Haustein, Costas, et al., 2015) and the Global South (Alperin, 2015). The focus on DOIs also represents a *de facto* reduction of altmetrics to journal articles, largely ignoring more diverse types of scholarly outputs, which, ironically, contradicts the diversification and democratization of scholarly reward that fuels the altmetrics movement. Above all, this repeats the frequently criticized biases of Web of Science and Scopus that it seeks to overcome.

Another major challenge of altmetrics presents itself in the strong dependency on data providers and aggregators. Similarly to the field of bibliometrics, which would not exist without the Institute for Scientific Information's Science Citation Index, the development of altmetrics is strongly shaped by data aggregators, particularly Altmetric. The loss of Altmetric would mean losing a unique data source[3]. Their collection of tweets, for instance, has become so valuable that another data aggregator (ImpactStory) decided to obtain data from them. This alludes to a monopolistic position that in the case of the Institute for Scientific Information has created the ubiquitous impact factor.

---

[3] The loss might be avoided or at least mitigated by maintaining a dark archive, which was mentioned by Altmetric founder Euan Adie in a tweet: https://twitter.com/stew/status/595527260817469440.

Even more challenging than the dependency on aggregators is the dependency on social media platforms as data providers. While a clear distinction can be made between papers' reference lists and the recorded citations in Scopus, Web of Science and Google Scholar, acts and recorded events are virtually identical on social media. If a citation database ceases to exist, it can still be reconstructed using the publications it was based on. If Twitter or Mendeley were discontinued, an entire data source would be lost and the acts of tweeting and saving to Mendeley would no longer exist[4], as they are not independent from the platforms themselves.

The strong dependency on social media platforms culminates in how their very nature directly affects user behavior and thus how technological affordances shape the actual acts. While this aspect has not been systematically analyzed, one could hypothesize that the use of Mendeley increases the number of cited references and the citation density in a document, that the tweet button increases the likelihood of tweeting, or that automated alerts, such as that of Twitter about trending topics, triggers further (re)tweets. The latter would represent a technology-induced Matthew effect for highly tweeted papers.

# 3 Conclusions and Outlook

For any metric to become a valid indicator of a social act, the act itself needs to be conceptualized (Lazarsfeld, 1993). The conceptualization can hence be used to construct meaningful indicators. In the case of altmetrics, recorded online events are used without having a proper understanding of underlying acts and in how far they are representative of various engagements with scholarly work. In fact, many of these acts are actually still forming and being shaped by technological affordances. Given the heterogeneity of the acts on which altmetrics are based, one would expect that a variety of concepts are needed to provide a clearer understanding of what is measured. Some recorded events might prove to be useful as filters, others might turn out to be valid measures of impact, but many might reflect nothing but buzz. In this context, it cannot be emphasized enough that social media activity does not equal social impact. Establishing a conceptual framework for scholarly metrics is particularly challenging when one considers that citations—which represent one single act—still lack a consensual theoretical framework after decades of being applied in research evaluation. While citations, despite being motivated by a number of reasons, have been and continue to be an essential part of the scholarly communication process, social media are just starting to enter academia and researchers are just beginning to incorporate them into their work routines.

The importance of data quality cannot be stressed enough particularly in today's *evaluation society* (Dahler-Larsen, 2012) and in a context, where any number beats no number. The dynamic nature of most of the events that altmetrics are based on provides a particular challenge with regards to their accuracy, consistency and reproducibility. Ensuring high data quality and sustainability is further impeded by the strong dependency on single data providers and aggregators. Above all, the majority of data is in the hands of for-profit companies, which contradicts the openness and transparency that has motivated the idea of altmetrics. As indicators will inevitably affect the processes which they intend to measure, adverse effects need to be mitigated by preventing misuse and avoiding to place too much emphasis on one indicator. While citation indicators have served as a great antagonist to help altmetrics gain momentum, it would not be in the interest of science to replace the impact factor by the Altmetric donut. It needs to be emphasized that any metric—be it citation or social media based—has to be chosen carefully with a view to the particular aim of the assessment exercise. The selection of indicators thus needs to be guided by the particular objectives of an evaluation, as well as by understanding their capabilities and constraints.

# Acknowledgements

The author acknowledges funding from the Alfred P. Sloan Foundation Grant # 2014-3-25 and would like to thank Vincent Larivière for stimulating discussions and helpful comments on the manuscript and Samantha Work for proofreading.

---

[4] In the short history of altmetrics, such loss can already be observed for Connotea.